\documentclass[reprint,aps,prb,twocolumn,superscriptaddress]{revtex4-1}
\usepackage{amssymb}
\usepackage[utf8]{inputenc}
\usepackage{amsmath}
\usepackage[]{graphicx}
\usepackage{bm}
\usepackage{sistyle}
\usepackage{color}
\usepackage{subfigure}

\begin{document}
\title{Analytical investigation of modulated spin torque oscillators in the framework of coupled differential equations with variable coefficients}

\author{Ezio Iacocca}
\email{ezio.iacocca@physics.gu.se}
\affiliation{Physics Department, University of Gothenburg, 412 96 G\"oteborg, Sweden}

\author{Johan \AA{}kerman}
\affiliation{Physics Department, University of Gothenburg, 412 96 G\"oteborg, Sweden}
\affiliation{Department of Microelectronics and Applied Physics, Royal Institute of Technology, Electrum 229, 164 40 Kista, Sweden}

\begin{abstract}
Modulation of Spin Torque Oscillators (STOs) is investigated by analytically solving the time-dependent coupled equations of an auto-oscillator. A Fourier series solution is proposed, leading to the coefficients being determined with a linear set of equations, from which a Nonlinear Amplitude and Frequency Modulation (NFAM) scheme is obtained. In this framework, the NFAM features are related to the intrinsic STO parameters, revealing a frequency-dependence of the harmonic-dependent modulation index that allows a modulation bandwidth to be defined for these devices. The presented results expose a rich parameter space, where the modulation and the STO's operation conditions define the observed modulation features. The Fourier-series representation of the time signal is suitable for studying periodic perturbations on the auto-oscillator equation.
\end{abstract}
\maketitle

\section{Introduction}

Spin torque oscillators~\cite{Silva2008} (STOs) are spin torque~\cite{Slonczewski1996,Berger1996,Ralph2008} driven devices whose high-frequency tunability and nanosized dimensions are promising for diverse technological applications. The oscillators' functionality can be very broad, depending on the specific application and its operation frequency. Two phenomena are of particular use, namely \emph{(i)} the ability to phase lock or synchronize to another periodic signal, and \emph{(ii)} the possibility of modulating a base-band signal.\cite{Carlson2002} In the case of STOs, both phenomena have been experimentally observed.

Phase locking of STOs has been extensively studied in the past few years. Injection locking of an STO to an external source is already well-established in both experimental~\cite{Rippard2005,Urazhdin2010,Quinsat2011} and numerical~\cite{Persson2007,Georges2008,Zhou2009,Serpico2009,Zhou2010,dAquino2010} works. Phase locking can also be achieved by mutually coupling several STOs. In this case, a distinction must be made between nanocontact~\cite{Rippard2004} and nanopillar~\cite{Tsoi2004} geometries. In nanocontacts, synchronization mediated by propagating spin waves~\cite{Slonczewski1999,Madami2011} has been satisfactorily achieved both experimentally~\cite{Mancoff2005,Kaka2005,Pufall2006} and numerically~\cite{Chen2009} for GMR-based STOs. In contrast, the mutual locking of nanopillars has been mainly successful in numerical studies.\cite{Grollier2006,Tiberkevich2009a,Iacocca2011}

Modulation of STOs has also been shown to be experimentally straightforward giving rise to several applications. STOs nanopillars have been proposed as hard-drive read-heads where the stray field from the perpendicular media acts a modulating source~\cite{Mizushima2011} both in GMR~\cite{Braganca2010} and TMR~\cite{Nagasawa2011} spin valves. From the perspective of communication applcations, Frequency Modulation (FM) has been observed in GMR nanocontacts~\cite{Pufall2005,Muduli2010,Muduli2011a,Muduli2011c,Pogoryelov2011,Pogoryelov2011a} and more recently in nano-oxide layer (NOL) nanocontact STOs.\cite{Mahdawi2011} Digital communication schemes, in particular Frequency Shift Keying (FSK), has been demonstrated in vortex-based nanocontact STOs up to the limit of analog FM.\cite{Manfrini2009,Manfrini2011} Similar modulation effects are most likely to be observed in other STO geometries that take advantage of Perpendicular Magnetic Anisotropy (PMA) materials~\cite{Mohseni2011} and nanopillars located in a microstrip resonator.\cite{Prokopenko2011}

In the pioneering frequency modulation experiment,\cite{Pufall2005} the authors took advantage of the strong phase--power coupling of GMR-based STOs to obtain a frequency-modulated voltage output from a base-band current tone. Although the gross features of frequency modulation were present, significant discrepancies from the model were observed:  \emph{(i)} an unexpected carrier frequency shift as a function of modulation strength, and \emph{(ii)} asymmetric sideband amplitudes in contrast to the expected \emph{n}-th order Bessel functions. A mathematical model was proposed by Consolo et al.~\cite{Consolo2010} to fit these features. Here, the temporal signal $s(t)$ was defined to be both amplitude-modulated and frequency-modulated
\begin{equation}
\label{eq:consolo}
    s(t) = A[x(t)] \cos(\omega_c + \omega_i[x(t)])t,
\end{equation}
where $\omega_c$ is the carrier frequency, and the amplitude $A$ and instantaneous frequency $\omega_i$ are polynomial expansions of a base-band message $x(t)$. This combined action of amplitude and frequency modulation was named ``Nonlinear Amplitude and Frequency Modulation'' (NFAM). Using this approximation, it was possible with very good accuracy to fit the GMR-STO modulation data with a polynomial expansion up to the third order.\cite{Muduli2010} However, the information acquired from these coefficients does not reflect the intrinsic mechanism of NFAM in STOs, nor its consequences for a specific choice of experimental conditions and device characteristics. To investigate the origin of NFAM in STOs, we here derive the modulation spectrum as a function of the intrinsic STO parameters.

The modulation spectrum of an STO is obtained by perturbing the auto-oscillator general equation~\cite{Slavin2009} with a slow time-varying tone. Such perturbation creates coupled phase and amplitude variations, which lead to NFAM. It is shown that a Fourier series gives an adequate representation of $s(t)$, in which case the problem can be reduced to determining the coefficients from a linear set of equations. Moreover, the Fourier coefficients obtained by this method give quantitative information, such as the carrier frequency shift, the modulation index, and the modulation bandwidth of the STO.

This paper is divided as follows: in Section II, the general formulation of the problem and its solution is given. It is shown that a carrier frequency shift appears as a consequence of power and phase coupling. Moreover, the modulation index is obtained by calculating the power spectral density, and shows an unexpected modulation-frequency dependence. In Section III, the analytical solution is evaluated for different modulation and operation conditions, and shows a qualitative agreement with experimental observations. Concluding remarks are given in Section IV.

\section{Modulation of a nonlinear auto-oscillator}
\begin{figure}[t!]
\centering
\includegraphics[scale=0.5]{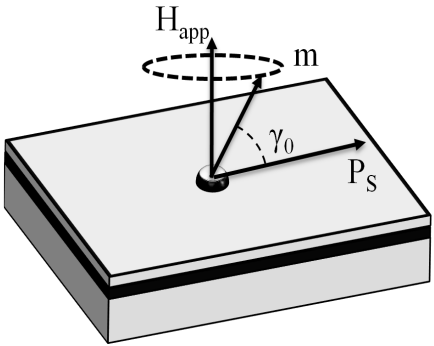}
\caption{\label{fig1} Nanocontact pseudo spin valve consisting of a thick (fixed) and a thin (free) magnetic layer (gray) decoupled by a thin metallic spacer (black). A nanocontact is patterned on top of the free layer, allowing the device to be driven with a spatially confined current. It is assumed that the auto-oscillator Eq.~(\ref{eq:slavin}) is a good approximation for such a structure. The free-layer magnetization $m$ is free to precess about the applied field $H_{app}$. The fixed layer is assumed to be magnetized predominantly in plane, and its magnetization $P_S$ subtends an angle $\gamma_0$ relative to the free layer magnetization. A steady precession is achieved when a spin polarized current exerts sufficient spin torque on the free layer to compensate for the magnetic damping.}
\end{figure}
In this section, the power spectral density (PSD) of a frequency-modulated nonlinear auto-oscillator is analytically calculated, using the general model proposed by Slavin and Tiberkevich:\cite{Slavin2009}
\begin{equation}
\label{eq:slavin}
    \frac{dc}{dt} + i\omega(p)c + \Gamma_+(p)c - \Gamma_-(p)c = 0,
\end{equation}
where $p$ is the oscillation's power, $c=\sqrt{p}e^{-i\phi}$ is the oscillation's amplitude, $\phi$ is its time-dependent phase, and $\omega(p)$, $\Gamma_+(p)$, and $\Gamma_-(p)$ are respectively the power-dependent oscillation frequency, damping, and negative damping. The power dependencies of these quantities are treated in Ref.~\onlinecite{Slavin2009} as polynomial expansions. We assume that the model of Eq.~(\ref{eq:slavin}) is a good description of the nanocontact geometry schematically shown in Fig.~\ref{fig1}. Here, a metallic nanocontact is patterned on top of an extended GMR spin valve. A strong magnetic field $H_{app}$ is applied to ensure the saturation of the magnetically active or ''free'' layer material. When a bias current is driven through the nanocontact, spin torque is exerted on the underlying free-layer area, counteracting the magnetic damping action, and leading to a small-angle precession about the applied field. Depending on the applied field angle, the free layer might exhibit propagating spin waves or a so-called localized bullet.\cite{Slavin2005,Gerhart2007,Consolo2007c,Bonetti2010} In this paper, we restrict our attention to a perpendicular applied field without loss of generality.

In order to solve the modulation problem, we recall the experiment of Ref.~\onlinecite{Pufall2005} where an ac current is added to the dc bias current. Consequently, we define the slow time-varying current as
\begin{equation}
\label{eq:fmcurrent}
    I(t) = I_{dc}\left(1 + \mu\cos{\omega_m t}\right),
\end{equation}
where $\mu$ is the modulation strength, defined as the ratio between the ac amplitude and the dc current, and $\omega_m$ is the modulation frequency. We assume that $\omega_m\ll\omega_{STO}$ (where $\omega_{STO}$ is the STO's free-running frequency) in order to be considered a modulating frequency. On the other hand, if $\omega_m\approx\omega_{STO}$, the STO might instead become injection-locked.\cite{} The current is thus included in the negative damping parameter, $\Gamma_-(p)$, which can be generally expanded as a polynomial in power
\begin{equation}
\label{eq:gammam}
    \Gamma_-(p) = \sigma I(t) \left(q_0 + q_1p + q_2p^2 + ... \right),
\end{equation}
where $\sigma = \epsilon \hbar \gamma / 2 e M_S V$, $\epsilon$ the spin-polarization efficiency, $\gamma$ the gyromagnetic ratio, $e$ the electron charge, $M_S$ the free layer's saturation magnetization, $V$ the free layer's current-carrying volume, and the coefficients $q_i$, $i=0,1,2,...$ are assumed to describe the sample's characteristics. It is further assumed that $I(t)$ creates a power perturbation of the form $p = p_0(1+2\delta p)$, where $p_0$ is the STO's free-running power. This approximation is valid as long as the modulation strength is small, which we consider a common scenario in STO modulation experiments (for instance, $\mu<0.1$ in both Ref.~\onlinecite{Pufall2005} and~\onlinecite{Muduli2010}).

Separating Eq.~(\ref{eq:slavin}) into real and imaginary parts, and expanding in power to first order, we obtain a set of differential equations with variable (time-dependent) coefficients for the STO's power and phase:
\begin{subequations}\label{eq:set}
\begin{eqnarray}
\label{eq:dpower}
    \frac{d\delta p}{dt} &=& \mu C_1\cos{\omega_mt}+\left(\mu C_2\cos{\omega_mt}-\Gamma_p\right)2\delta p,\\
\label{eq:dphase}
    \frac{d\phi}{dt} &=& \omega_{STO} + 2\nu\Gamma_p\delta p,
\end{eqnarray}
\end{subequations}
where we define the constants
\begin{subequations}\label{eq:constants}
\begin{eqnarray}
\label{eq:c1}
    C_1 &=& \Gamma_-(p_0),\\
\label{eq:c2}
    C_2 &=& \left(\Gamma_-(p_0)+\frac{d\Gamma_-(p)}{dp} \Big |_{p_0}p_0\right).
\end{eqnarray}
\end{subequations}

Here, $\omega_{STO}=\omega_o + \nu\Gamma_p$, $\omega_o = \gamma\mu_0(H-M_S)$ is the FMR frequency, $\mu_0$ is the vacuum permeability, $\Gamma_p = \alpha\omega_o(\xi-1)$ is the restoration rate, $\alpha$ is the Gilbert damping parameter, $\xi=I_{dc}/I_{th}$ is the supercriticality parameter, and $I_{th}$ is the threshold current for spin-torque-driven oscillations. The restoration rate is assumed to be constant with respect to the free-running oscillations. Qualitatively, it gives a measure of how fast the STO reacts to a perturbation, and so plays a fundamental role determining the synchronization speed for these devices.\cite{Zhou2010,dAquino2010,Iacocca2011}

We propose a Fourier series as a solution of Eq.~(\ref{eq:set})a, based on the periodicity of the time-dependent term:
\begin{equation}
\label{eq:Fouriersol}
    \delta p = A_0 + \sum\limits_{n=1}^\infty {A_n\sin{n\omega_mt}+B_n\cos{n\omega_mt}}.
\end{equation}

Introducing this solution into Eq.~(\ref{eq:set})a gives an infinite system of equations for the Fourier coefficients. It is possible to further reduce the problem to the determination of $A_n$, leading to a system of linear equations which can be solved numerically (see Appendix A). The solution, expressed as a single sinusoid, is
\begin{subequations}\label{eq:solution}
\begin{eqnarray}
\label{eq:power}
    \delta p &=& A_0 + \sum\limits_{n=1}^\infty {\sqrt{B_n^2+A_n^2}\cos(n\omega_mt-\psi_n)},\\
    \phi &=& (\omega_{STO}+2\nu\Gamma_pA_0)t \nonumber \\
\label{eq:phase}
    &+& \sum\limits_{n=1}^\infty {\frac{2\nu\Gamma_p}{n\omega_m}\sqrt{B_n^2+A_n^2}\sin(n\omega_mt-\psi_n)}.
\end{eqnarray}
\end{subequations}

The notation of Eq.~(\ref{eq:solution}) explicitly shows a harmonic-dependent phase $\psi_n=\arctan(A_n/B_n)$. This phase plays a fundamental role in the form of the PSD, as discussed below. The first indication of NFAM is visible in the first right-hand term of Eq.~(\ref{eq:phase}), where the oscillation frequency is shifted by $2\pi f_s = 2\nu\Gamma_pA_0$. From the exact solution given in Eq.~(\ref{eq:ACoeff}), we obtain
\begin{equation}
\label{eq:fs}
    2\pi f_s = -\frac{\mu \nu C_2B_1}{2}.
\end{equation}

Consequently, the shift is directly proportional to the modulation strength and to the STO nonlinearities, via the constant $C_2$.

In order to identify the next feature of NFAM---namely the asymmetry of the sideband---the PSD must be calculated. By expressing the sinusoidal functions as exponentials and performing Taylor expansion (Appendix B), the PSD can be expressed as a series of convolutions
\begin{eqnarray}
\label{eq:PSD}
    PSD &=& p_0\delta(\omega_{STO'})*\Big | \Big [ (1+A_0)\delta(0)\nonumber \\
    &+&\sum\limits_{n=1}^\infty {\frac{\bar{X}_n}{2}\delta(n\omega_m)+\frac{X_n}{2}\delta(-n\omega_m)} \Big ]\nonumber \\
    &*&\prod\limits_{n=1}^\infty J_0(\beta_n)\delta(0)+\sum\limits_{j=1}^\infty \frac{J_j(\beta_n)}{|X_n|^j}(\bar{X}_n^j\delta(nj\omega_m)\nonumber \\
    &+&(-1)^jX_n^j\delta(-nj\omega_m))\Big |
\end{eqnarray}
where we define the complex variable $X_n=B_n+iA_n$ (hence $\psi_n = $~arg$(X_n)$), the notation $\bar{(\cdot)}$ represents the complex conjugate, and the shifted carrier frequency is written as $\omega_{STO'} = \omega_{STO}+2\pi f_s$. The notation $\delta(x_0) = \delta(x-x_0)$ is used, where $\delta$ is the Kronecker delta. The product symbol represents in this case a series of convolutions. The real harmonic-dependent modulation index is defined as
\begin{equation}
\label{eq:beta}
    \beta_n = \frac{2\nu\Gamma_p|X_n|}{n\omega_m}.
\end{equation}

Comparison with the FM modulation index, $h=\Delta f/f_m$, suggests that the peak frequency deviation for STOs can be defined as $2\pi\Delta f_n = 2\nu\Gamma_p|X_n|$ for given harmonic frequency and modulation conditions. Due to the form of $|X_n|$, it is expected to be linearly dependent on the modulation strength $\mu$ and proportional to $[(n\omega_m)^2 + (2\Gamma_p)^2]^{-1/2}$. The latter dependence is of particular interest in terms of the modulation bandwidth (MBW) discussed in Section III.

Several features of the spectrum can be readily identified from the right-hand side of Eq.~(\ref{eq:PSD}). First we can see the shifted carrier frequency which, by virtue of the properties of convolution, shifts the base-band spectrum in frequency. Second, a nonlinear amplitude-modulation (NAM) spectrum arises from the power fluctuations. Third, a series of FM spectra whose harmonics expand with $n$ can be seen to arise from the combined contributions of the phase modulation and the nonlinearly enhanced power fluctuations. In the case of a weakly nonlinear oscillator ($\nu \leq 1$), the FM series of convolutions can be neglected, and the spectrum reduces to a NAM spectrum. For a particular applied field angle, the condition $\nu =0$ is satisfied, and the spectrum reduces to pure AM.\cite{Consolo2010a} On the other hand, if the time-dependent term is neglected ($C_2\approx 0$), only the first harmonic coefficients of Eq.~(\ref{eq:ACoeff}) will be non-zero, and the PSD will take the form of pure FM.\cite{Slavin2009}

The convoluting terms in Eq.~(\ref{eq:PSD}) define the power of each harmonic. Although it is tedious to obtain a meaningful analytical expression, it can be inferred that the power of the harmonics is sensitive to the phase $\psi_n$. We can illustrate this by approximating each convoluting term to the \emph{second} harmonic, and solving for the first upper and lower sideband power. By expanding the absolute values to first order, we obtain the power difference of the sideband $\Delta = 2J_1(\beta_1)(1+A_0)p_0$, which is in general non-vanishing. This crude approximation does not represent the asymmetry of Eq.~(\ref{eq:PSD}), which is affected up to the fifth harmonic, as shown in Appendix A.

Summarizing this section, we have obtained a fairly complex spectrum by solving the set of Eq.~(\ref{eq:set}). This complexity, along with the interdependence of several key parameters such as the modulation strength and frequency, limits the analytical insight that can be gained. However, the frequency shift of the carrier can be explicitly obtained by considering a Fourier-series approach, and the sideband asymmetry arises as a consequence of the solution's harmonic-dependent phases. The solution of the spectrum leads to the definition of the harmonic-dependent modulation index, from which the STO peak frequency deviation can be defined.

\section{Numerical results}

\begin{figure}[t!]
\centering
\includegraphics[scale=0.5]{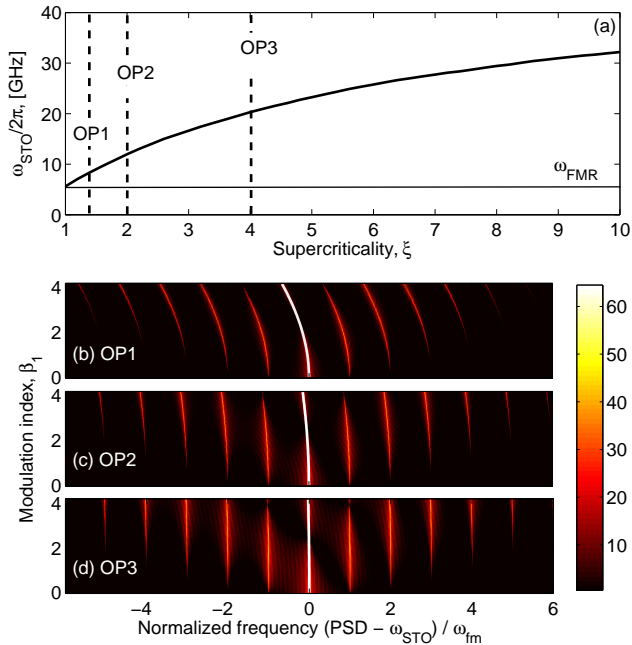}
\caption{\label{fig2}(Color online) (a) Frequency dispersion of the STO considered in the numerical calculations, as a function of the supercriticality $\xi$. Beginning from $\xi=1$, current-driven oscillations are sustained. The frequency increases nonlinearly from the FMR frequency indicated by the horizontal line. Three operation points, OP1, OP2, and OP3, are selected for the following results. (b-d) PSD as a function of the modulation index $\beta_1$ and normalized frequency scale. The frequency shift is proportional to both the modulation strength and the curvature of (a), following Eq.~(\ref{eq:fs}) (solid white line over the carrier).}
\end{figure}

In this section, we consider a first order expansion of the STOs parameters in order to evaluate the PSD of Eq.~(\ref{eq:PSD}). The coefficients are calculated up to the tenth harmonic by solving the linear system of equations Eq.~(\ref{eq:ACoeff}) in matrix form. We consider the nanocontact geometry of Fig.~\ref{fig1} with the parameters $\mu_0H_{app}=1$~T, $\mu_0M_S=0.8$~T, $\gamma=28$~GHz/T, $\alpha=0.01$, and $\nu=100$. Evaluating the STO parameters $\Gamma_+(p)$, $\Gamma_-(p)$, and $\omega(p)$ according to auto-oscillator theory,\cite{Slavin2009} we obtain the frequency vs supercriticality dispersion shown in Fig.~\ref{fig2}a. As expected, the frequency is equal to the FMR frequency at the threshold of the oscillation ($\xi=1$).

Three operation points (OP), OP1, OP2, and OP3, were selected in Fig.~\ref{fig2}a, such that their restoration rates were $\Gamma_p/2\pi=11.2$, $44.8$, and $156.8$~MHz, respectively. Each OP was modulated, and their PSDs, calculated as the FFT of Eq.~(\ref{eq:BPSD}), are shown in Fig.~\ref{fig2}(b-d) as a function of $\beta_1$ and a fixed modulation frequency, $f_m = 100$~MHz. The first-harmonic modulation index is used as a reference, since it is assumed to be the main contribution to the sideband power. In order to evaluate Eq.~(\ref{eq:ACoeff}), the modulation strength is calculated from $\beta_1$, returning different ranges for each OP. The frequency axis is normalized to the modulation frequency and centered on the carrier frequency. On this scale, all operation points share the same number of visible sidebands, as well as the position of the carrier. It is observed that the frequency shift is more pronounced near the threshold (OP1). This is expected from Eq.~(\ref{eq:fs}), shown as the white lines in Fig.~\ref{fig2}(b-d), since the modulation strength is higher near both the threshold and the curvature of Fig~\ref{fig2}(a). Such curvature-dependence has been observed in experiments~\cite{Muduli2010} away from the oscillation threshold, presumably due to higher order nonlinearities. These effects can be included in the present framework through Eq.~(\ref{eq:c2}), as the derivative of the negative damping may be substantially different. A numerical fit of experimental data is not within the scope of the present paper, but we argue that such a fit could be achieved by measuring the auto-oscillator parameters, as has recently been shown in Ref.~\onlinecite{Urazhdin2010}.
\begin{figure}[t!]
\centering
\includegraphics[scale=0.5]{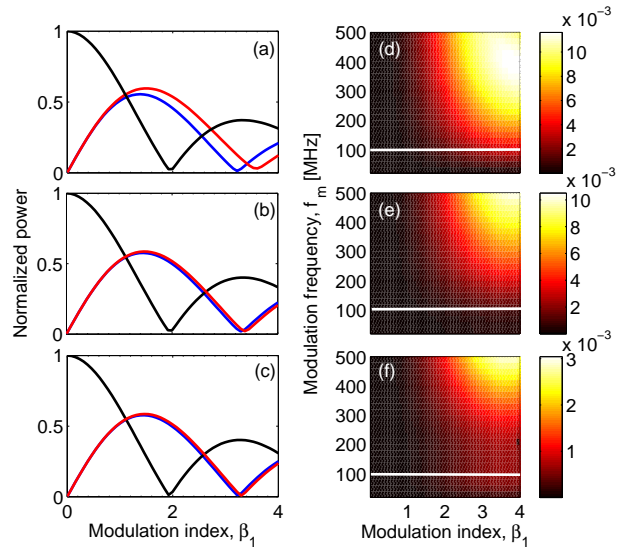}
\caption{\label{fig3}(Color online) (a-c) Carrier (black) and first upper (red) and lower (blue) sidebands for each OP as a function of the modulation index $\beta_1$. The sideband asymmetry is clearly visible, and is substantially higher at OP1 due to the strong frequency shift. Panels (d-f) show the normalized power difference of the sideband $\Delta$ as a function of both modulation strength and frequency. White lines denote the slice shown in (a-c). The colormaps show that the asymmetry increases together with both modulation parameters.}
\end{figure}

The power of the normalized carrier and of the first upper and lower sidebands is shown in Fig.~\ref{fig3}(a-c) for each OP and with modulation frequency $f_m = 100$~MHz. The combined action of the frequency shift and the harmonic-dependent phase introduced by the Fourier series results in power asymmetry between the sidebands. This asymmetry is notably higher for OP1 in correlation with its similarly enhanced frequency shift. In the color plots of Fig.~\ref{fig3}(d-f), the power difference of the first upper and lower sidebands $\Delta$ is shown as a function of $\beta_1$ and $f_m$. The white horizontal line denotes the slice shown in the panels (a-c), respectively. Being a common feature, the sideband asymmetry increases together with both the modulation strength and the frequency. These dependencies can be qualitatively understood from Eq.~(\ref{eq:PSD}) and Eq.~(\ref{eq:ACoeff}). A stronger $\mu$ increases the power of the higher harmonic coefficients, so that they become non-negligible in the PSD. Thus, an increase in the asymmetry is expected. On the other hand, the dependence on the modulation frequency arises from the relation between $A_n$ and $B_n$. Roughly, one can approximate $A_1 = \omega_m/2\Gamma_p B_1$, so that the complex variables in the PSD can be expressed as $X_1 = B_1 (1+i\omega_m/2\Gamma_p)$. From here, it is clear that as $\omega_m$ increases past the parameter $2\Gamma_p$, the imaginary term becomes dominant or, in other words, the harmonic phase increases towards $\pm \pi/2$. Consequently, the power contribution of each harmonic to the sideband becomes heavily weighted by the phase, leading to enhanced asymmetry.

\begin{figure}[t!]
\centering
\includegraphics[scale=0.5]{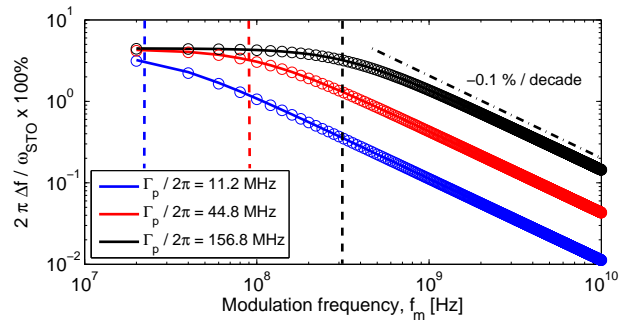}
\caption{\label{fig4}(Color online) Peak frequency deviation $\Delta f$ dependence on the modulation frequency. The solid lines have been approximately calculated from the first-order modulation index, while the circles are have been calculated from the maximum amplitude of the instantaneous frequency. The dashed lines represent the cut-off frequency $2\Gamma_p$ for each OP while the dash-dotted black line is a guide for the eye, showing a 0.1~\%/decade slope.}
\end{figure}

Finally, we discuss the features of the peak frequency deviation $\Delta f_n$ in relation to the modulation bandwidth. The frequency dependence of $\Delta f$ for each OP and fixed $\mu=0.05$ is shown in Fig.~\ref{fig4}, both by evaluating the first-harmonic modulation index, Eq.~(\ref{eq:beta}) (solid lines), and by calculating the maximum of the instantaneous frequency from Eq.~(\ref{eq:phase}) (circles). The magnitude of the deviation is shown relative to the OP oscillator frequency, while the logarithmic scale is used to enhance the frequency dependence. Both approaches agree very well, so that the first harmonic modulation index suffices to perform the following calculations approximately. A practical consequence of this feature is that choosing a maximum modulation strength does not guarantee the frequency excursion estimated from the STO's frequency vs current characteristics. This is closely related to the modulation bandwidth of oscillators. 

The modulation bandwidth (MBW) gives a measure of the frequency range in which an oscillator has optimal modulation properties. A common criterion is the 3~dB power attenuation that is likewise used to characterize filters. Since the power in a FM scheme depends on the Bessel functions, the methods used to measure the MBW rely on indirectly estimating the degradation of the linearity of the modulation index. In the present framework, the modulation index is defined from the calculation of the power spectrum (Appendix B), and one can directly estimate the MBW from Eq.~(\ref{eq:beta}). This can be accomplished by considering a vanishing modulation frequency and an arbitrary $\mu$, so that the first-harmonic modulation index is $\beta_i$. The task is then to keep the ratio $\mu / \omega_m$ constant while looking for $\beta_f = \beta_i / \sqrt{2}$. Performing this calculation gives the result that the MBW for STOs is $2\Gamma_p$. Since the lack of linearity in our framework is given by $2\pi\Delta f$, one can now understand Fig.~\ref{fig4} as a STO transfer function under modulation having a characteristic low-pass filter form with a 0.1~\%/decade roll-off slope (dash-dotted black line) after the cut-off frequency $2\Gamma_p$ (dashed vertical lines). 

\section{Conclusions}

The NFAM spectrum is obtained from the auto-oscillator general model by considering the power perturbation created by the modulating signal. These power fluctuations are enhanced by the STO's nonlinearity, resulting in a spectrum consisting of a series of convolutions between NAM and FM spectra. Its implicit dependence on the total damping parameter and nonlinearities suggests that the NFAM characteristics will exhibit sample-to-sample variation. In fact, such characteristics have been observed in experiments,\cite{Muduli2010}  and their dependence on the modulation parameters qualitatively agrees with the results shown here. In contrast to the originally proposed model,\cite{Consolo2010} we found a frequency dependence of the modulation index which leads to the estimation of the modulation bandwidth for STOs. Moreover, the STO under modulation has a low-pass filter behavior with a cut-off frequency given by $2\Gamma_p$, which usually depends on the operation point of a specific sample. Although the modulation bandwidth for STOs has yet to be measured experimentally, we expect a correlated degradation of STO modulation characteristics in frequency-dependent studies. On the other hand, the Fourier series solution proposed in this paper can, in principle, reduce to a linear set of coupled equations any STO geometry described by the auto-oscillator general equation Eq.~(\ref{eq:slavin}) perturbed by slow time-varying signal.

Support from the Swedish Research Council (VR) is gratefully acknowledged. Johan~\r{A}kerman is a Royal Swedish Academy of Sciences Research Fellow supported by a grant from the Knut and Alice Wallenberg Foundation.

\appendix
\section{}

The Fourier series solution proposed in Eq.~(\ref{eq:Fouriersol}) is introduced into Eq.~(\ref{eq:dpower}). The variable term can be expanded using trigonometric identities:
\begin{eqnarray}
\label{eq:Aexpansion}
    \cos(\omega_m t)\delta p &=& A_0\cos(\omega_m t) \\
    &+&\sum\limits_{n=1}^\infty\frac{A_n}{2}[\sin(\omega_m(1+n)t)-\sin(\omega_m(1-n)t)]\nonumber \\
    &+&\frac{B_n}{2}[\cos(\omega_m(1+n)t)+\cos(\omega_m(1-n)t)].\nonumber
\end{eqnarray}

This series can be then rearranged by expanding the summation and identifying equal harmonics. Changing the index accordingly, we obtain
\begin{eqnarray}
\label{eq:Aexpansion2}
    \cos(\omega_m t)\delta p &=& (A_0 + \frac{B_1}{2}) \\
    &+& \frac{1}{2} [(2B_2 A_0)\cos(\omega_m t) - A_2\sin(\omega_m t)] \nonumber \\
    &+& \frac{1}{2} \sum\limits_{n=2}^\infty (A_{n-1}-A_{n+1})\sin(n\omega_m t)\nonumber \\
    &+& (B_{n-1}+B_{n+1})\sin(n\omega_m t).\nonumber
\end{eqnarray}

This expression treats the harmonics separately, so that Eq.~(\ref{eq:dpower}) can be solved by collecting harmonic terms. The resulting system of equations for the coefficients is
\begin{subequations}\label{eq:ACoeff}
\begin{eqnarray}
\label{eq:AC1}
    0 &=& \mu\frac{C_2}{2}B_1 + 2\Gamma_p A_0,\\
\label{eq:AC2}
    \omega_m A_1 &=& \mu (C_1 - C_2 A_0 - \frac{C_2}{2}B_2) - 2\Gamma_p B_1,\\
\label{eq:AC3}
    \omega_m B_1 &=& 2\Gamma_p A_1 - \mu\frac{C_2}{2} A_2,\\
\label{eq:AC4}
    0 &=& \sum\limits_{n=2}^\infty 2\Gamma_p B_n + n\omega_mA_n\\
    && \; \; \; +\mu\frac{C_2}{2}(B_{n-1}-B_{n+1}),\nonumber \\
\label{eq:AC5}
    0 &=& \sum\limits_{n=2}^\infty 2\Gamma_p A_n - n\omega_mB_n\\
    && \; \; \; + \mu\frac{C_2}{2}(A_{n-1}-A_{n+1}).\nonumber
\end{eqnarray}
\end{subequations}

\begin{figure}[t!]
\centering
\includegraphics[scale=0.45]{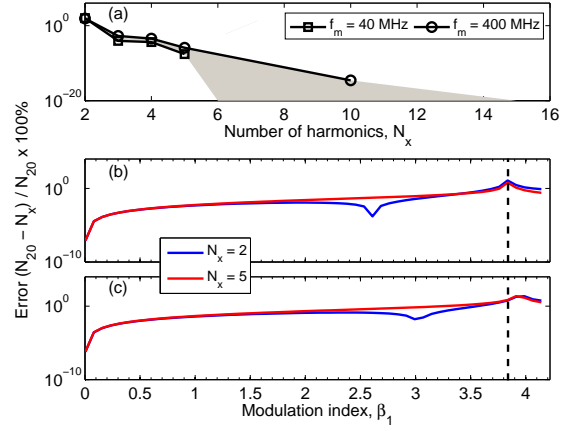}
\caption{\label{fig:error}(Color online) (a) Total error as a function of the number of harmonics $N$ in the series, calculated from the linear equations Eq.~(\ref{eq:ACoeff}). For $N\geq5$, the error is assumed to be negligible. Two modulation frequencies are shown as limiting cases. For intermediate frequencies, the error lies in the shaded area. (b-c) Error between the linear set of equations and the recursive method as a function of the modulation index $\beta_1$. The modulation frequencies for each case are $40$ and $400$~MHz. The error introduced by the recursive method is generally small, but diverges close to the minimum of the sideband, which is indicated by vertical dashed lines.}
\end{figure}

There are two possible solutions to the set of Eq.~(\ref{eq:ACoeff}). One can, for instance, express a system of equations depending only on the $A$s, so that $A_n\propto A_{n+2},A_{n+1},A_{n-1},A_{n-2}$. This new system can easily be solved numerically using matrix algebra. The error will depend on the size of the matrix or, in other words, on the maximum harmonic where the series is truncated. The total error is shown in Fig.~\ref{fig:error}(a) for two frequencies of different orders of magnitude. The case for $N=20$ harmonics is used as a reference. It is observed that in both cases, $N\geq5$ returns a total error smaller than  $10^{-5}$~\%.

Another solution is grounded on the fact that the harmonic coefficients decay with $n$, so that we can assume in general that $A_n\propto A_{n-1},A_{n-2}$. This approach can be implemented numerically as a recurrent series, and this is shown for different values of $N$ and $f_m$ in Fig.~\ref{fig:error}(b-c). Here we observe that $N=5$ also converges to a minimum error in this approximation. However, as $\beta_1$ is swept out, it is clear the the error increases towards the minimum of the first sideband (indicated by a dashed line). It is noteworthy that the maximum error in Fig.~\ref{fig:error}(c) is slightly shifted, which corresponds to the impact of higher harmonic terms.

\section{}

The power spectral density (PSD) of the proposed solution is obtained from Eq.~(\ref{eq:solution}), which defines both time-dependent power and phase variations. The PSD is defined as
\begin{equation}
\label{eq:BPSD}
    PSD = p_0 * \Big | \hat{f}(1+\delta p) * \hat{f}(\cos(\phi)) \Big |,
\end{equation}
where $\hat{f}$ is the Fourier transform. In the following we perform each Fourier transform separately, and express the result as a convolution. The first term on the right hand side has the form of a NAM. The phase introduced by the sine function leads us to define the complex variable $X_n=B_n + iA_n$, so that
\begin{eqnarray}
\label{eq:BAM}
    \hat{f}(1+\delta p) &=& (1+A_0)\delta(0)\\
    && \; \; +\sum\limits_{n=1}^\infty {\frac{\bar{X}_n}{2}\delta(n\omega_m)+\frac{X_n}{2}\delta(-n\omega_m)},\nonumber
\end{eqnarray}
where the notation $\delta(x_0) = \delta(x-x_0)$ is used and $\delta$ is the Kronecker delta.

The second term on the right hand side of Eq.~(\ref{eq:BPSD}) is calculated by using Euler's formulae, and subsequently expanding in Taylor series. We obtain
\begin{eqnarray}\nonumber
\label{eq:Bcosphi}
    \cos(\phi) &=& \frac{e^{-i\omega_{STO'}}}{2} \prod\limits_{n=1}^\infty \Big ( \sum\limits_{k=0}^\infty \frac{(b_n-ia_n)^k}{k!2^k}e^{ink\omega_mt}\\
    &\cdot& \sum\limits_{k=0}^\infty (-1)^k\frac{(b_n+ia_n)^k}{k!2^k}e^{ink\omega_mt} \Big ) + c.c,
\end{eqnarray}
where $a_n = 2\nu\Gamma_p A_n/(n\omega_m)$, $b_n = 2\nu\Gamma_p B_n/(n\omega_m)$ and $c.c$ is the complex conjugate. The two summations can be multiplied by expanding and rearranging the terms. Defining $2x_n=b_n+ia_n$, it becomes possible to rewrite the product terms as
\begin{eqnarray}
\label{eq:Bsumprod}
    \prod\limits_{n=1}^\infty(\cdot) &=& \sum\limits_{k=0}^\infty (-1)^k\frac{x_n\bar{x_n}}{k!k!}\\
    &+& \sum\limits_{k=0}^\infty \sum\limits_{j=1}^\infty (-1)^k\frac{x_n\bar{x_n}}{k!(k+j)!}\left[ (\bar{x_n}e)^j + (-x_ne^{-1})^j \right ]. \nonumber
\end{eqnarray}

From this expression, we can identify the summations on $k$ as Bessel functions of order $j$ and argument $\beta_n=2\sqrt{x_n\bar{x_n}}$ (upon renormalization of the coefficients by $\beta_n/2$). The product takes the form
\begin{eqnarray}
\label{eq:Bsumprod2}
    \prod\limits_{n=1}^\infty(\cdot) &=& J_0(\beta_n)\nonumber \\
    &+& \sum\limits_{j=1}^\infty \frac{2J_j(\beta_n)}{\beta_n} \left[ (\bar{x_n}e)^j + (-x_ne^{-1})^j \right ].
\end{eqnarray}

The Fourier transform of Eq.~(\ref{eq:Bsumprod2}) is easily performed by expressing the result as a product of convolutions. By further evaluating $x_n$ in terms of the complex variable $X_n$, and convoluting each harmonic contribution, we obtain Eq.~(\ref{eq:PSD}).

\bibliographystyle{aipnum4-1}
\end{document}